\documentstyle [12pt,epsfig]{article}
\title{
Crystal experiments on efficient beam extraction
       }
\author{
A.G.Afonin, \underbar{V.M.Biryukov}, V.T.Baranov, V.N.Chepegin,\\
Y.Chesnokov, V.Kotov, V.Terekhov,
E.Troyanov \\
{\em\small IHEP-Protvino, Russia}\\
V.Guidi, G.Martinelli, M.Stefancich, D.Vincenzi,
{\em\small Ferrara Un., Italy}; \\
Yu.Ivanov,  {\em\small PNPI, Russia};
D.Trbojevic,  {\em\small BNL, USA} \\
W.Scandale,  {\em\small CERN};
M.Breese,  {\em\small Surrey Un., UK}
}
\date{Presented at {\em Int. Conf. on Atom. Coll. in Solids} \\
(ICACS-19: Paris, July 29 - August 3, 2001)}
\begin{document}
\maketitle

\begin{abstract}
Silicon crystal was channeling and extracting 70-GeV protons from the U-70
accelerator with efficiency of 85.3$\pm$2.8\% as measured
for a beam of $\sim$10$^{12}$ protons directed towards crystals
of $\sim$2 mm length in spills of $\sim$2 s duration.
The experimental data follow very well the
prediction of Monte Carlo simulations.
This success is important to devise a more efficient use of the
U-70 accelerator in Protvino and
provides a crucial support for
implementation of crystal-assisted collimation
of gold ion beam in RHIC and slow extraction from AGS onto E952,
now in preparation at Brookhaven Nat'l Lab.
Future applications, spanning in the energy from sub-GeV (medical) to
order of 1 GeV (scraping in the SNS, extraction from COSY) to order of 1 TeV
and beyond (scraping in the Tevatron, LHC, VLHC), can benefit from these
studies.
\end{abstract}

The technique of bent crystal channeling to steer particle beams,
with applications in extraction from accelerator and in beam
collimation, has progressed rapidly thanks to efforts at
IHEP Protvino\cite{book}, CERN\cite{akbari},
and FNAL\cite{fnal}.
Beams of up to 10$^6$ proton/s were extracted from CERN SPS and Tevatron
by Si crystals of just 4 cm in length, with typical efficiencies
on the order of 10-20\%.
It was predicted\cite{theory}
that  efficiency of crystal channeling extraction can be boosted
to much higher values
by multiple particle encounters with a shorter crystal
installed in a circulating beam.
To clarify this mechanism a new experiment was started at IHEP
at the end of 1997,
with intention to test very short crystals and achieve very high
efficiencies of extraction\cite{pac99,epac2000}.

The benefits of a crystal-assisted extraction are fourfold.
In hadron colliders this mode of extraction can in general be made compatible
with the colliding mode of operation.
The time structure of the extracted beam is practically flat,
since the extraction mechanism is resonance-free.
The size of the extracted beam is smaller and more round than
in a resonant extraction. Finally polarized beams can be extracted without
detrimental effects on the polarization.

The classic two-stage collimation system for loss localisation
in accelerators typically uses a small scattering target as a primary
element and a bulk absorber as secondary element\cite{jeann}.
The role of the primary element is to give a substantial angular kick
to the incoming particles in order to increase the impact parameter
on the secondary element, which is generally placed in the optimum position
to intercept transverse or longitudinal beam halo.
An amorphous primary target scatters the impinging particles
in all possible directions.
Ideally, one would prefer to use
a "smart target" which kicks all particles in
only one direction: for instance, only in radial direction, only outward,
and only into the preferred angular range corresponding to the centre of
the absorber (to exclude escapes).
A bent crystal is the practical implementation for such a smart target:
it traps particles and conveys them into the desired direction
\cite{near}.

These two possible applications of crystal channeling in modern hadron
accelerators,  extraction and halo collimation,
can be exploited in a broad range of energies,
from sub-GeV cases (i.e. for medical accelerators) to multi-TeV machines
(for high-energy research). Indeed,
several projects are in progress to investigate them.
Crystal collimation is studied
at RHIC (100-250 GeV)\cite{rhic},
considered for the Tevatron (1000 GeV)\cite{tev}
and the LHC\cite{lhc},
for the Spallation Neutron Source (1 GeV)\cite{nuria},
whilst
crystal-assisted slow extraction is considered
for AGS (25 GeV)\cite{woody}
and COSY (1-2 GeV).
In all cases, the critical issue is the channeling efficiency.

In the last two years, we demonstrated crystal channeling
with 50\% efficiency\cite{pac99,epac2000}.
We also showed that, when properly aligned, these crystals could
be efficiently used as primary collimators, thereby reducing
by a factor two the
radiation level measured downstream of the collimation region of U-70.
To continue our investigations,
we installed and tested several new crystals in different
straight sections of the U-70 ring.
Three of them were produced by different manufacturers with a new shape.
They were made with narrow strips of Si(111),
about 40 mm long vertically and a fraction of mm
long in the radial direction. Their azimuthal length was only a few mm.
They were bent by a metallic holder providing
deflections of 0.8 to 1.5 mrad.

\begin{figure}[h]
\begin{center}
\parbox[c]{10.5cm}{\epsfig{file=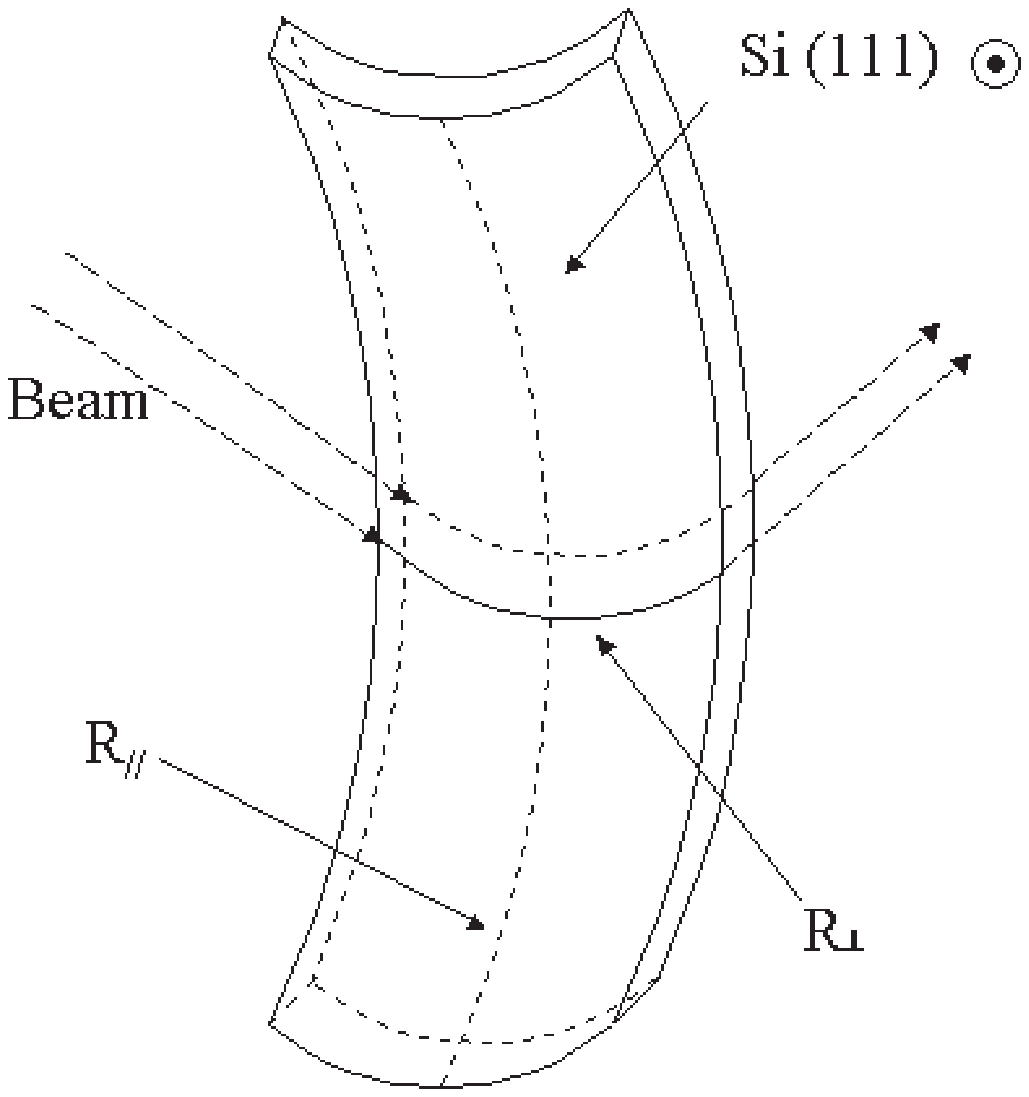,width=11 cm }}
\end{center}
        \caption{
Sketch of a crystal deflector
shaped as a saddle due to curvature in both
vertical and radial directions. This design was used to obtain 85\%
efficiency of extraction.
} \label{crystal}
        \end{figure}

To make a crystal deflector this short
(factor of $\sim$20 shorter than the crystals used at CERN and FNAL),
a new design has been developed, Fig.\ref{crystal}.
The crystal strip has the shape of a saddle, being bent in both
vertical and radial directions.
The advantages of "new-generation" crystals are threefold:
(a) they can be made shorter along the beam than in previous designs,
(b) they have no straight ends as the bending mechanism is continuous,
and (c) they have no amorphous material close to the beam
(like the "legs" of U- and O-shaped deflectors used earlier in CERN and IHEP).

Two crystals were assembled in Protvino:
one $2$ mm long was bent by 0.9 mrad, the other $4$ mm
long was bent by 1.5 mrad.
The third crystal, $1.8$ mm long, 0.8 mrad bent, was prepared at the
University of Ferrara  and
chemically polished at optical level to remove the defects induced
during diamond slicing.
The two Russian crystals were used in extraction mode,
whilst the Italian one was
tested as a primary collimator.
The three crystals were exposed to 70 GeV proton beams and
used to channel and extract halo particles.

Figure \ref{col2} illustrates the beneficial effect of crystals
when used as primary collimators.
It shows beam profiles in the radial direction
downstream of the crystal
as measured on the entry face of the collimator.
Four cases are reported.
First, an amorphous collimator is used as primary target
whilst the crystal is kept outside of the beam envelope.
As expected, the beam profile is peaked at the collimator edge,
Fig.\ref{col2}(a).
In the second case, Fig.\ref{col2}(b), the
crystal is used as the primary scraper but is not aligned to the beam.
When properly aligned, Fig.\ref{col2}(c), the crystal channels most
of the incoming beam into the depth of the collimator.
In the last case, Fig.\ref{col2}(d),
the beam is simply kicked by a magnet towards the secondary collimator,
whilst the crystal is retracted.

The extraction efficiency is given by
the ratio of the extracted beam intensity as measured in
the external beam line to all the beam loss as measured in the entire ring.
We obtained very high efficiencies in each of the
three new crystals: namely, both the 1.8 and 2 mm long crystals reached
85\% efficiency, whilst the 4mm long crystal
reached 68\% efficiency.
These striking results were obtained in a steady manner
over many runs.

In Fig.\ref{mcs4} we plot the predicted\cite{epac2000}
(by Monte Carlo code\cite{pre51}) and the measured extraction
efficiencies together with the data obtained earlier.
The agreement between measurements and simulations is excellent.

Beside the channeling efficiency,
the ability to withstand a high beam intensity and the crystal lifetime
also important for a practical application.
Crystals located in the region upstream of the U-70 cleaning area were
irradiated with the entire circulating beam, spilt out in a rather
short time durations to simulate very dense halo collimation.
Analysis has shown that our crystals were irradiated up to
$2 \times 10^{14}$ particles per spill of $\sim$1 s duration.
When averaged over machine cycles, the irradiation rate was as high as
$2 \times 10^{13}$ proton hits/s.
Notice that this irradiation rate
compares with the expected beam
loss rate at the Spallation Neutron Source. Indeed,
the SNS Accumulator Ring should generate a 1 GeV
proton flux
of 60$\times$2$\times$10$^{14}$ per second. At the expected rate of
beam loss of 0.1--1\% the halo flux will be
(1.2--12)$\times$10$^{13}$ protons/s.
Crystals may well tolerate
high flux of protons
as large as those expected to hit the SNS beam collimation system.

Another crystal, $5$ mm long,
was exposed for several minutes to even higher radiation flux.
Beams of up to 10$^{13}$ 70-GeV protons (resulting in $\sim$10$^{14}$
proton hits) were directed towards the crystal
in spills of 50 ms every 9.6 s.
After this extreme exposure,
the channeling properties of the crystal were tested
in an external beam. The deflected beam observed with photo-emulsion
\cite{heacc} was perfectly normal, without breaks, nor significant
tails eventually produced by dechanneled particles.
Several crystals in use in U-70 have been exposed to high intensity beams
for months, thereby accumulating very large integrated doses\cite{epac2000}.
After the estimated irradiation of $\sim$10$^{20}$p/cm$^2$
the initial channeling efficiency of about 43\% was practically unaffected.
This irradiation is still below
the world highest results\cite{bnl,cern}.
The CERN experiment\cite{cern} showed that at
5$\times$10$^{20}$ p/cm$^2$ the crystal lost only 30\% of its
deflection efficiency, which means $\sim$100 years lifetime
in the intense beam of NA48 experiment.

\begin{figure}[h]
\begin{center}
\setlength{\unitlength}{.75mm}
\begin{picture}(105,140)(20,-40)
\thicklines

\put(0,0) {\line(1,0){150}}
\put(0,0) {\line(0,1){120}}
\put(0,120) {\line(1,0){150}}
\put(150,0){\line(0,1){120}}

\multiput(0,0)(10,0){15}{\line(0,1){3}}
\multiput(0,120)(10,0){15}{\line(0,-1){3}}
\multiput(0,0)(0,25){5}{\line(1,0){3}}
\multiput(150,0)(0,20){7}{\line(-1,0){3}}

\put(0,-9){\makebox(1,1)[b]{0}}
\put(20,-10){\makebox(1,1)[b]{2.5}}
\put(40,-10){\makebox(1,1)[b]{5}}
\put(60,-10){\makebox(1,1)[b]{7.5}}
\put(80,-10){\makebox(1,1)[b]{10}}
\put(100,-10){\makebox(1,1)[b]{12.5}}
\put(-17,75){\makebox(1,.5)[l]{300}}
\put(-17,50){\makebox(1,.5)[l]{200}}
\put(-17,25){\makebox(1,.5)[l]{100}}

\put(-10,115){\large $I$}
\put(125,-10){$R$(mm)}
\put(100,107){\Large\bf 1.3 GeV}

\linethickness{.3mm}
\put(100,0.9)  {\line(1,0){10}}
\put( 90,7.5)  {\line(1,0){10}}
\put( 80,18)  {\line(1,0){10}}
\put( 70,37)  {\line(1,0){10}}
\put( 60,50)  {\line(1,0){10}}
\put( 50,43.2)  {\line(1,0){10}}
\put( 40,32.7)  {\line(1,0){10}}
\put( 30,25.5)  {\line(1,0){10}}
\put( 20,24.7)  {\line(1,0){10}}
\put( 10,28.3)  {\line(1,0){10}}
\put( 0,41)  {\line(1,0){10}}

\put( 110,0.9)  {\line(0,-1){0.9}}
\put( 100,7.5)  {\line(0,-1){7.5}}
\put( 90,18)  {\line(0,-1){18}}
\put( 80,37)  {\line(0,-1){37}}
\put( 70,50)  {\line(0,-1){50}}
\put( 60,50)  {\line(0,-1){50}}
\put( 50,43.2)  {\line(0,-1){43.2}}
\put( 40,32.7)  {\line(0,-1){32.7}}
\put( 30,25.5)  {\line(0,-1){25.5}}
\put( 20,28.3)  {\line(0,-1){28.3}}
\put( 10,41)  {\line(0,-1){41}}

\linethickness{7mm}
\put( 80,0)  {\line(0,1){14}  }
\put( 70,0)  {\line(0,1){33.2}}
\put( 60,0)  {\line(0,1){45.6}}
\put( 50,0)  {\line(0,1){17.2}}
\put( 40,0)  {\line(0,1){6.7} }
\end{picture}

\end{center}
\caption{
Beam profile as measured on the collimator entry face
with 1.3 GeV protons. In black is shown the simulated
profile of channeled protons.
}
  \label{colg}
\end{figure}
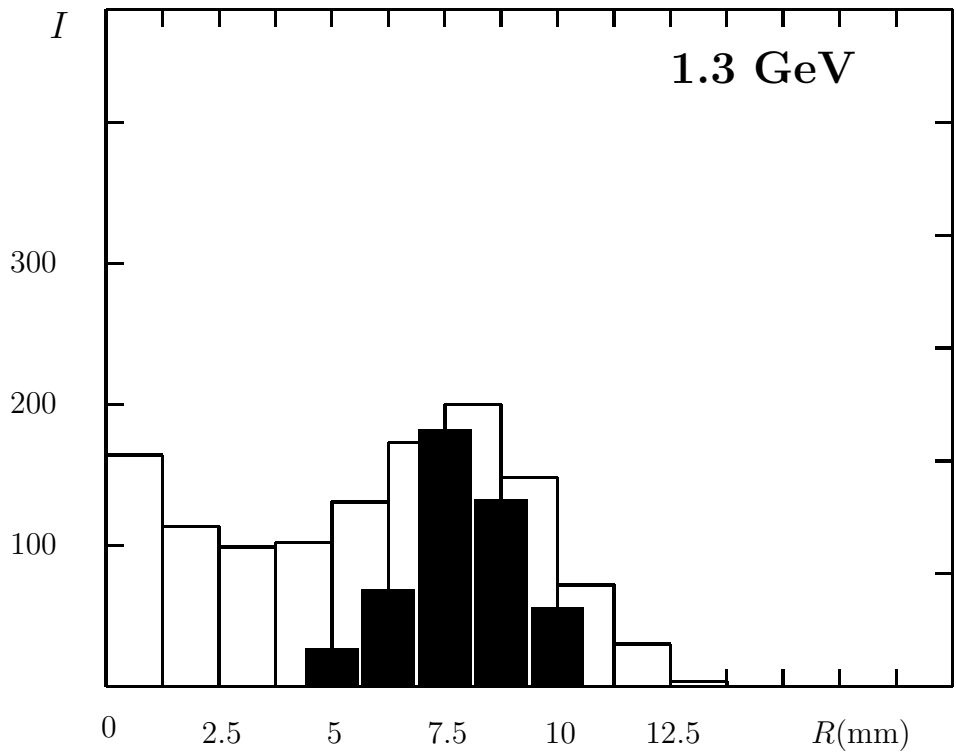

On the same location in U-70 ring
with the same 1.8-mm crystal of Si(111)
positioned $\sim$20 m upstream of the ring collimator,
we have repeated the crystal collimation experiment
at the injection flattop of U-70,
proton kinetic energy of 1.3 GeV.
With the crystal aligned to the incoming halo particles,
the radial beam profile at the collimator entry face
showed a significant channeled peak far from the edge, Fig.\ref{colg}.
About half of the protons intercepted by the collimator jaw,
have been channeled there by a crystal; i.e., the crystal has
doubled the amount of particles intercepted by the jaw.
As only part (about 34\%) of all particles scattered off the crystal
have reached the jaw, we estimate the
crystal deflection efficiency as 15-20\%.
The observed
figure of efficiency could be well reproduced in computer simulations.
This figure is orders of magnitude higher than previous world data
for low-GeV energy range.
It is remarkable that the same crystal was efficiently channeling
and really helping in cleaning halo particles both at 70 GeV and at 1.3 GeV,
thus demonstrating to be operational in a very wide energy range.

In summary,
the crystal channeling efficiency has reached unprecedented high values
both at top energy and at injection energy.
The same 2 mm long crystal was used to channel 70 GeV protons
with an efficiency of 85.3$\pm$2.8\% during several weeks of operation
and 1.32 GeV protons with an efficiency of 15-20\% during some test runs.
Crystals with a similar design were able to stand radiation doses
over $10^{20}$ proton/cm$^2$
and irradiation rates of $2 \times 10^{14}$ particles
incident on crystal in spills of $\sim$2 s duration
without deterioration of their performances.

The efficiency results well match the figures theoretically expected
for ideal crystals. As simulations show, extraction
and collimation with channeling efficiencies over 90-95\% is feasible.
The obtained high figures provide
a crucial support for the ideas to apply this technique in beam cleaning
systems, for instance in RHIC and Tevatron.
Earlier Tevatron scraping simulations\cite{tev} have shown that a crystal
scraper can reduce accelerator-related backgrounds in CDF and D0
experiments by a factor of $\sim$10.
This year, first experimental data is expected from RHIC
where a crystal collimator\cite{rhic} is installed.
The technique presented here is potentially applicable also in LHC for
instance to improve the efficiency of the LHC cleaning system by embedding
bent crystals in the primary collimators\cite{lhc}.
This work was supported by INTAS-CERN grant 132-2000,
RFBR grant 01-02-16229, and by the
"Young researcher Project" of the University of Ferrara.

\clearpage

\begin{figure}[h]
\begin{center}
\setlength{\unitlength}{.35mm}
\begin{picture}(105,120)(20,-40)
\thicklines

\put(0,0) {\line(1,0){150}}
\put(0,0) {\line(0,1){120}}
\put(0,120) {\line(1,0){150}}
\put(150,0){\line(0,1){120}}

\multiput(0,0)(10,0){15}{\line(0,1){3}}
\multiput(0,120)(10,0){15}{\line(0,-1){3}}
\multiput(0,0)(0,25){5}{\line(1,0){3}}
\multiput(150,0)(0,25){5}{\line(-1,0){3}}

\put(0,-13){\makebox(1,1)[b]{0}}
\put(20,-13){\makebox(1,1)[b]{2.5}}
\put(40,-13){\makebox(1,1)[b]{5}}
\put(60,-13){\makebox(1,1)[b]{7.5}}
\put(80,-13){\makebox(1,1)[b]{10}}
\put(100,-13){\makebox(1,1)[b]{12.5}}
\put(-12,100){\makebox(1,.5)[l]{4}}
\put(-12,75){\makebox(1,.5)[l]{3}}
\put(-12,50){\makebox(1,.5)[l]{2}}
\put(-12,25){\makebox(1,.5)[l]{1}}

\put(-10,115){\large $I$}
\put(115,-13){$R$(mm)}
\put(124,100){\large\bf (a)}

\linethickness{.3mm}
\put( 70,2.5)  {\line(1,0){10}}
\put( 60,1.5)  {\line(1,0){10}}
\put( 50,5)  {\line(1,0){10}}
\put( 40,7)  {\line(1,0){10}}
\put( 30,9)  {\line(1,0){10}}
\put( 20,12)  {\line(1,0){10}}
\put( 10,16)  {\line(1,0){10}}
\put( 0,27.5)  {\line(1,0){10}}
\put( 80,2.5)  {\line(0,-1){2.5}}
\put( 70,2.5)  {\line(0,-1){2.5}}
\put( 70,1.5)  {\line(0,-1){1.5}}
\put( 60,5)  {\line(0,-1){5}}
\put( 50,7)  {\line(0,-1){7}}
\put( 40,9)  {\line(0,-1){9}}
\put( 30,12)  {\line(0,-1){12}}
\put( 20,16)  {\line(0,-1){16}}
\put( 10,27.5)  {\line(0,-1){27.5}}
\end{picture}

\begin{picture}(105,140)(20,-40)
\thicklines

\put(0,0) {\line(1,0){150}}
\put(0,0) {\line(0,1){120}}
\put(0,120) {\line(1,0){150}}
\put(150,0){\line(0,1){120}}

\multiput(0,0)(10,0){15}{\line(0,1){3}}
\multiput(0,120)(10,0){15}{\line(0,-1){3}}
\multiput(0,0)(0,25){5}{\line(1,0){3}}
\multiput(150,0)(0,25){5}{\line(-1,0){3}}

\put(0,-13){\makebox(1,1)[b]{0}}
\put(20,-13){\makebox(1,1)[b]{2.5}}
\put(40,-13){\makebox(1,1)[b]{5}}
\put(60,-13){\makebox(1,1)[b]{7.5}}
\put(80,-13){\makebox(1,1)[b]{10}}
\put(100,-13){\makebox(1,1)[b]{12.5}}
\put(-12,100){\makebox(1,.5)[l]{4}}
\put(-12,75){\makebox(1,.5)[l]{3}}
\put(-12,50){\makebox(1,.5)[l]{2}}
\put(-12,25){\makebox(1,.5)[l]{1}}

\put(-10,115){\large $I$}
\put(115,-13){$R$(mm)}
\put(124,100){\large\bf (b)}

\linethickness{.3mm}
\put( 50,.5)  {\line(1,0){10}}
\put( 40,2.5)  {\line(1,0){10}}
\put( 30,4)  {\line(1,0){10}}
\put( 20,6.5)  {\line(1,0){10}}
\put( 10,10.5)  {\line(1,0){10}}
\put( 0,25.5)  {\line(1,0){10}}
\put( 60,.5)  {\line(0,-1){.5}}
\put( 50,2.5)  {\line(0,-1){2.5}}
\put( 40,4)  {\line(0,-1){4}}
\put( 30,6.5)  {\line(0,-1){6.5}}
\put( 20,10.5)  {\line(0,-1){10.5}}
\put( 10,25.5)  {\line(0,-1){25.5}}
\end{picture}

\begin{picture}(105,140)(20,-40)
\thicklines

\put(0,0) {\line(1,0){150}}
\put(0,0) {\line(0,1){120}}
\put(0,120) {\line(1,0){150}}
\put(150,0){\line(0,1){120}}

\multiput(0,0)(10,0){15}{\line(0,1){3}}
\multiput(0,120)(10,0){15}{\line(0,-1){3}}
\multiput(0,0)(0,25){5}{\line(1,0){3}}
\multiput(150,0)(0,25){5}{\line(-1,0){3}}

\put(0,-13){\makebox(1,1)[b]{0}}
\put(20,-13){\makebox(1,1)[b]{2.5}}
\put(40,-13){\makebox(1,1)[b]{5}}
\put(60,-13){\makebox(1,1)[b]{7.5}}
\put(80,-13){\makebox(1,1)[b]{10}}
\put(100,-13){\makebox(1,1)[b]{12.5}}
\put(-12,100){\makebox(1,.5)[l]{4}}
\put(-12,75){\makebox(1,.5)[l]{3}}
\put(-12,50){\makebox(1,.5)[l]{2}}
\put(-12,25){\makebox(1,.5)[l]{1}}

\put(-10,115){\large $I$}
\put(115,-13){$R$(mm)}
\put(124,100){\large\bf (c)}

\linethickness{.3mm}
\put(80,99)  {\line(1,0){10}}
\put(70,56)  {\line(1,0){10}}
\put( 60,9.6)  {\line(1,0){10}}
\put( 50,0)  {\line(1,0){10}}
\put( 40,0)  {\line(1,0){10}}
\put( 30,2.4)  {\line(1,0){10}}
\put( 20,2)  {\line(1,0){10}}
\put( 10,3.6)  {\line(1,0){10}}
\put( 0,8)  {\line(1,0){10}}
\put( 90,99)  {\line(0,-1){99}}
\put( 80,99)  {\line(0,-1){99}}
\put( 70,56)  {\line(0,-1){56}}
\put( 60,9.6)  {\line(0,-1){9.6}}
\put( 40,2.4)  {\line(0,-1){2.4}}
\put( 30,2.4)  {\line(0,-1){2.4}}
\put( 20,3.6)  {\line(0,-1){3.6}}
\put( 10,8)  {\line(0,-1){8}}
\end{picture}

\begin{picture}(105,110)(20,-10)
\thicklines

\put(0,0) {\line(1,0){150}}
\put(0,0) {\line(0,1){120}}
\put(0,120) {\line(1,0){150}}
\put(150,0){\line(0,1){120}}

\multiput(0,0)(10,0){15}{\line(0,1){3}}
\multiput(0,120)(10,0){15}{\line(0,-1){3}}
\multiput(0,0)(0,25){5}{\line(1,0){3}}
\multiput(150,0)(0,25){5}{\line(-1,0){3}}

\put(0,-13){\makebox(1,1)[b]{0}}
\put(20,-13){\makebox(1,1)[b]{2.5}}
\put(40,-13){\makebox(1,1)[b]{5}}
\put(60,-13){\makebox(1,1)[b]{7.5}}
\put(80,-13){\makebox(1,1)[b]{10}}
\put(100,-13){\makebox(1,1)[b]{12.5}}
\put(-12,100){\makebox(1,.5)[l]{4}}
\put(-12,75){\makebox(1,.5)[l]{3}}
\put(-12,50){\makebox(1,.5)[l]{2}}
\put(-12,25){\makebox(1,.5)[l]{1}}

\put(-10,115){\large $I$}
\put(115,-13){$R$(mm)}
\put(124,100){\large\bf (d)}

\linethickness{.3mm}
\put(120,14.5)  {\line(1,0){10}}
\put(110,35)  {\line(1,0){10}}
\put(100,44)  {\line(1,0){10}}
\put( 90,40)  {\line(1,0){10}}
\put( 80,20)  {\line(1,0){10}}
\put( 70,5)  {\line(1,0){10}}
\put(130,14.5)  {\line(0,-1){14.5}}
\put(120,35)  {\line(0,-1){35}}
\put(110,44)  {\line(0,-1){44}}
\put(100,44)  {\line(0,-1){44}}
\put( 90,40)  {\line(0,-1){40}}
\put( 80,20)  {\line(0,-1){20}}
\put( 70,5)  {\line(0,-1){5}}
\end{picture}

\end{center}
\caption{
Beam profiles measured on the collimator entry face:
(a) crystal is out, beam scraped by collimator alone;
(b) crystal is in the beam, but misaligned;
(c) crystal is in the beam, aligned;
(d) crystal is out, beam kicked by magnet.
}
  \label{col2}
\end{figure}
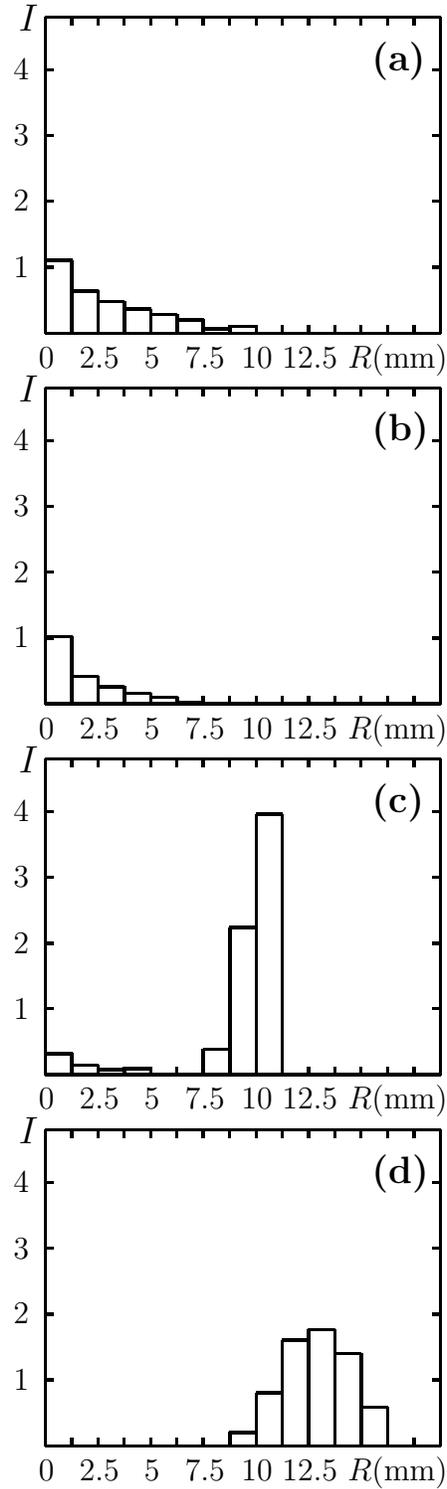

\begin{figure}[thb]
\begin{center}
\setlength{\unitlength}{.8mm}
\begin{picture}(110,100)(15,-10)
\thicklines
\linethickness{.3mm}
\put(   8,91)  {\circle{2.5}}
\put(  10,92)  {\circle{2.5}}
\put(  12,92)  {\circle{2.5}}
\put(  14,92)  {\circle{2.5}}
\put(  16,92)  {\circle{2.5}}
\put(  20,90)  {\circle{2.5}}
\put(  24,89)  {\circle{2.5}}
\put(  30,87)  {\circle{2.5}}
\put(  40,83)  {\circle{2.5}}
\put(  50,79)  {\circle{2.5}}
\put(  60,75)  {\circle{2.5}}

\put(  78,67)  {\LARGE $\star$}
\put(  34,84)  {\LARGE $\star$}
\put(  38,84)  {\LARGE $\star$}

\put( 138,12)  {$\otimes$}
\put(  58,62)  {$\Box$}
\put(  98,48)  {$\Box$}

\put(89,91) {(o)  ideal strip, predicted}
\put(89,84) {($\star$, $\otimes$)  strips, measured}
\put(89,77) {($\Box$)  O-shape, measured}

\put(0,0) {\line(1,0){150}}
\put(0,0) {\line(0,1){100}}
\put(0,100) {\line(1,0){150}}
\put(150,0){\line(0,1){100}}

\multiput(4,0)(4,0){37}{\line(0,1){2}}
\multiput(20,0)(20,0){7}{\line(0,1){4}}
\multiput(4,100)(4,0){37}{\line(0,-1){2}}
\multiput(20,100)(20,0){7}{\line(0,-1){4}}
\multiput(0,20)(0,20){4}{\line(1,0){3}}
\multiput(0,4)(0,4){25}{\line(1,0){1.4}}
\multiput(150,20)(0,20){4}{\line(-1,0){3}}
\multiput(150,4)(0,4){25}{\line(-1,0){1.4}}

\put(0,-7){\makebox(1,1)[b]{0}}
\put(20,-7){\makebox(1,1)[b]{1}}
\put(40,-7){\makebox(1,1)[b]{2}}
\put(60,-7){\makebox(1,1)[b]{3}}
\put(80,-7){\makebox(1,1)[b]{4}}
\put(100,-7){\makebox(1,1)[b]{5}}
\put(120,-7){\makebox(1,1)[b]{6}}
\put(140,-7){\makebox(1,1)[b]{7}}
\put(-11,20){\makebox(1,.5)[l]{20}}
\put(-11,40){\makebox(1,.5)[l]{40}}
\put(-11,60){\makebox(1,.5)[l]{60}}
\put(-11,80){\makebox(1,.5)[l]{80}}
\put(-13,100){\makebox(1,.5)[l]{100}}

\put(10,105){\Large Extraction (channeling) efficiency (\%)}
\put(60,-14){\Large Crystal length (mm)}

\end{picture}
\end{center}
\caption{
Crystal extraction efficiency as measured for 70-GeV protons.
Recent results ($\star$, strips 1.8, 2.0, and 4 mm),
results of 1999-2000 ($\Box$, O-shaped crystals 3 and 5 mm),
and of 1997 ($\otimes$, strip 7 mm).
Also shown (o) is Monte Carlo prediction [8]
for a perfect crystal with 0.9 mrad bending.
}
  \label{mcs4}
\end{figure}
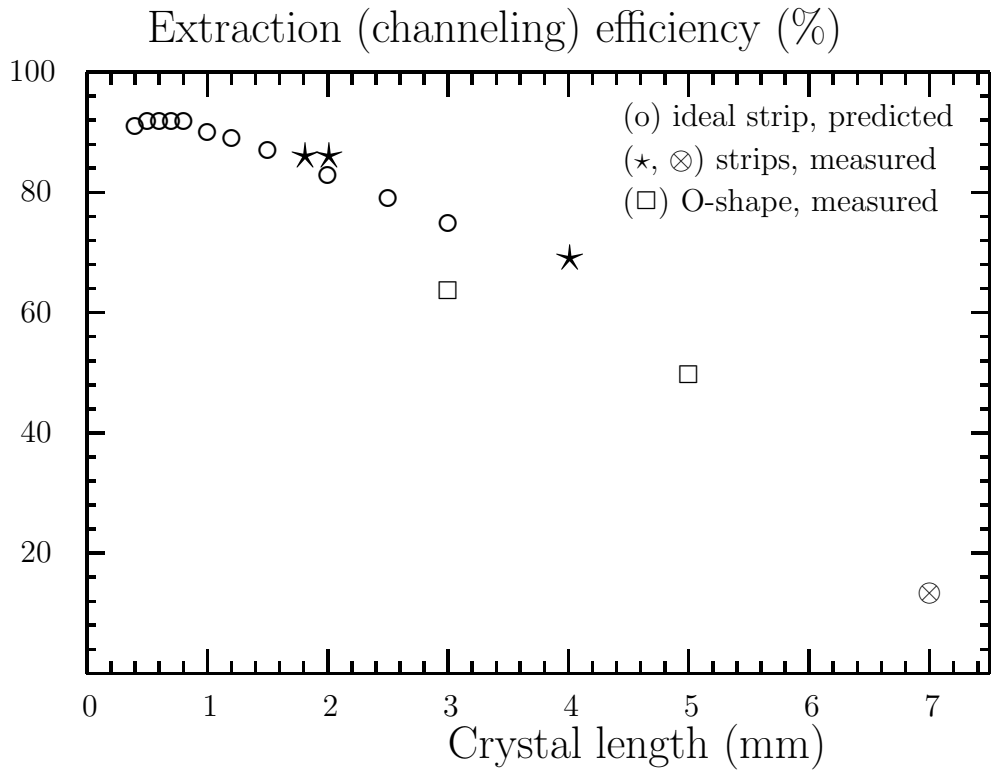

\begin{figure}[h]
\begin{center}
\setlength{\unitlength}{.75mm}
\begin{picture}(110,140)(20,-10)
\thicklines
\linethickness{.2mm}

\put(0,0) {\line(1,0){150}}
\put(0,0) {\line(0,1){120}}
\put(0,120) {\line(1,0){150}}
\put(150,0){\line(0,1){120}}

\multiput(0,0)(5,0){30}{\line(0,1){3}}
\multiput(10,0)(15,0){9}{\line(0,1){5}}
\multiput(0,120)(5,0){30}{\line(0,-1){3}}
\multiput(10,120)(15,0){9}{\line(0,-1){5}}
\multiput(0,50)(0,50){2}{\line(1,0){5}}
\multiput(0,10)(0,10){11}{\line(1,0){3}}
\multiput(150,50)(0,50){2}{\line(-1,0){5}}
\multiput(150,10)(0,10){11}{\line(-1,0){3}}

\put(10,-7){\makebox(1,1)[b]{-12}}
\put(25,-7){\makebox(1,1)[b]{-9}}
\put(40,-7){\makebox(1,1)[b]{-6}}
\put(55,-7){\makebox(1,1)[b]{-3}}
\put(70,-7){\makebox(1,1)[b]{0}}
\put(85,-7){\makebox(1,1)[b]{3}}
\put(100,-7){\makebox(1,1)[b]{6}}
\put(115,-7){\makebox(1,1)[b]{9}}
\put(130,-7){\makebox(1,1)[b]{12}}
\put(8,100){\makebox(1,.5)[l]{1}}
\put(8,50){\makebox(1,.5)[l]{0.5}}
\put(138,100){\makebox(1,.5)[l]{1}}
\put(132,50){\makebox(1,.5)[l]{0.5}}

\put(2,126){\Large Intensity (10$^{14}$ particles per spill)}
\put(30,-18){\Large Position on crystal, $H$(mm)}

\linethickness{1.mm}
\put(  57.5,103)  {\line(1,0){25}}

\linethickness{.3mm}
\put(  57.5,81.3)  {\line(1,0){25}}
\put(  57.5,81.3)  {\line(0,-1){42}}
\put(  82.5,81.3)  {\line(0,-1){42}}
\put(  32.5,39.3)  {\line(1,0){25}}
\put(  82.5,39.3)  {\line(1,0){25}}
\put( 107.5,13  )  {\line(1,0){25}}
\put( 132.5, 3.5)  {\line(1,0){17.5}}
\put(   7.5,13  )  {\line(1,0){25}}
\put(     0, 3.5)  {\line(1,0){7.5}}
\put( 107.5,39.3)  {\line(0,-1){25.7}}
\put(  32.5,39.3)  {\line(0,-1){25.7}}
\put( 132.5, 3.5)  {\line(0,1){9.5}}
\put(   7.5,13  )  {\line(0,-1){9.5}}

\end{picture}
\end{center}
        \caption{
        Intensity of proton hits at the crystal per spill
        of $\sim$1 s duration.
        This figure takes into account the number of hits on crystal per
        incident proton ($\sim$100 for short crystal).
        The thick line at the top shows the intensity achieved at
        the 5-mm-high O-shaped crystal.
        The thin line below shows the intensity achieved at
        the 40-mm-high "strip" crystal.
}
        \end{figure}
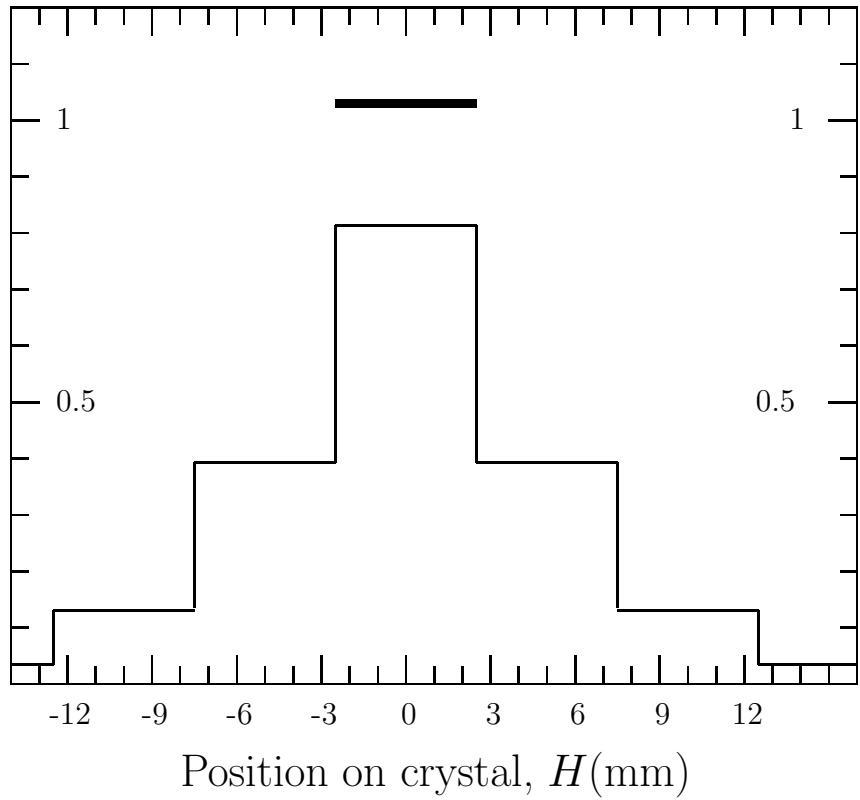

\begin{figure}[h]
\begin{center}
\setlength{\unitlength}{.75mm}
\begin{picture}(110,160)(20,-10)
\thicklines
\linethickness{.2mm}

\put(0,0) {\line(1,0){150}}
\put(0,0) {\line(0,1){100}}
\put(0,100) {\line(1,0){150}}
\put(150,0){\line(0,1){100}}

\multiput(25,0)(4,0){25}{\line(0,1){2}}
\multiput(25,0)(20,0){6}{\line(0,1){4}}
\multiput(25,100)(4,0){25}{\line(0,-1){2}}
\multiput(25,100)(20,0){6}{\line(0,-1){4}}
\multiput(0,30)(0,30){3}{\line(1,0){3}}
\multiput(0,6)(0,6){16}{\line(1,0){1.4}}
\multiput(150,30)(0,30){3}{\line(-1,0){3}}
\multiput(150,6)(0,6){16}{\line(-1,0){1.4}}

\put(25,-7){\makebox(1,1)[b]{0}}
\put(45,-7){\makebox(1,1)[b]{0.1}}
\put(65,-7){\makebox(1,1)[b]{0.2}}
\put(85,-7){\makebox(1,1)[b]{0.3}}
\put(105,-7){\makebox(1,1)[b]{0.4}}
\put(125,-7){\makebox(1,1)[b]{0.5}}
\put(-6,30){\makebox(1,.5)[l]{1}}
\put(-6,60){\makebox(1,.5)[l]{2}}
\put(-6,90){\makebox(1,.5)[l]{3}}
\put(153,30){\makebox(1,.5)[l]{1}}
\put(153,60){\makebox(1,.5)[l]{2}}
\put(153,90){\makebox(1,.5)[l]{3}}

\put(2,106){\Large Integrated Dose (10$^{20}$particle/cm$^2$)}
\put(40,-18){\Large Position in crystal, R(mm)}

\linethickness{1.mm}
\put(  25,64.2)  {\line(1,0){20}}
\put(  45,64.2)  {\line(0,-1){27.6}}
\put(  45,36.6)  {\line(1,0){20}}
\put(  65,36.6)  {\line(0,-1){6.3}}
\put(  65,30.3)  {\line(1,0){20}}
\put(  85,30.3)  {\line(0,-1){4.6}}
\put(  85,25.7)  {\line(1,0){20}}
\put( 105,25.7)  {\line(0,-1){2.5}}
\put( 105,23.2)  {\line(1,0){20}}

\linethickness{.3mm}
\put(  25,94.2)  {\line(1,0){20}}
\put(  45,94.2)  {\line(0,-1){27.2}}
\put(  45,67.)  {\line(1,0){20}}
\put(  65,67.)  {\line(0,-1){10}}
\put(  65,57.)  {\line(1,0){20}}
\put(  85,57.)  {\line(0,-1){7.8}}
\put(  85,49.2)  {\line(1,0){20}}
\put( 105,49.2)  {\line(0,-1){1.5}}
\put( 105,47.7)  {\line(1,0){20}}

\linethickness{.4mm}
\multiput(  25,26.8)(4,0){5}{\line(1,0){2}}
\multiput(  45,26.8)(0,-4){4}{\line(0,-1){2}}
\multiput(  45,9.3)(4,0){5}{\line(1,0){2}}
\multiput(  65,9.3)(0,-0.9){2}{\line(0,-1){.45}}
\multiput(  65,7.5)(4,0){5}{\line(1,0){2}}
\multiput(  85,6.3)(4,0){5}{\line(1,0){2}}
\multiput( 105,6.)(4,0){5}{\line(1,0){2}}

\end{picture}
\end{center}
\caption{
Irradiation of the crystal entry face (O-shaped crystal)
in proton hits/cm$^2$,
after 10$^5$ machine cycles ($\sim$1 month of accelerator run)
with dump of 10$^{12}$ proton/cycle.
Shown for extraction efficiency 43\% (thick line, middle);
for misaligned crystal (thin line, top);
for extraction efficiency 73\% (dashed, bottom).
}
  \label{mcs3}
\end{figure}
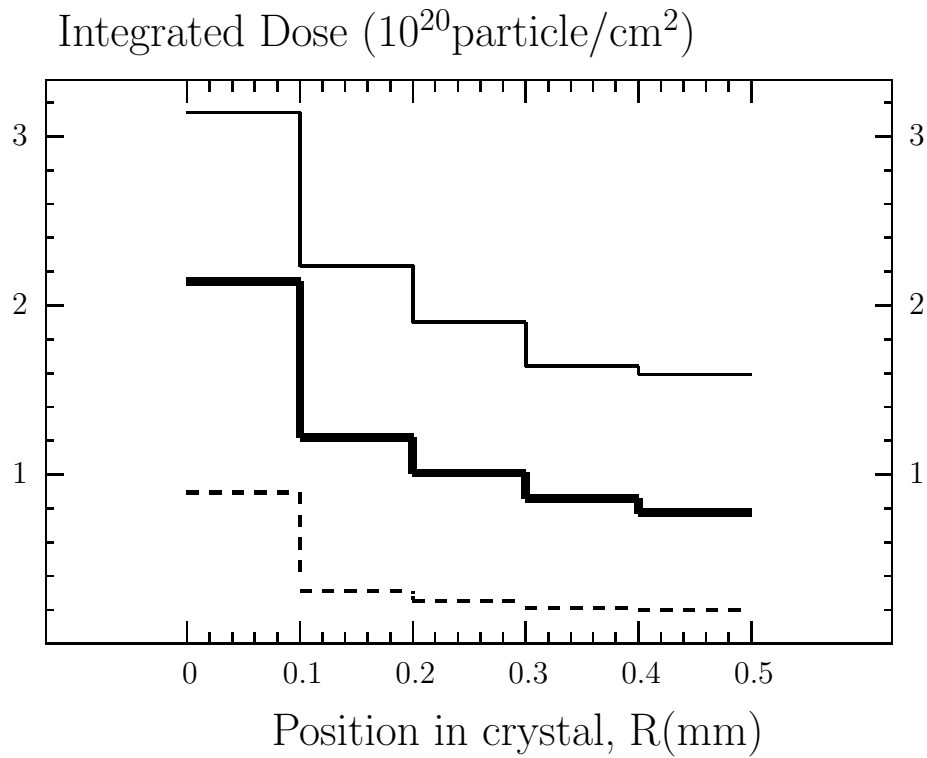

\end{document}